\pdfoutput=1

\documentclass[letterpaper,twocolumn,10pt]{article}
\usepackage{usenix2019,epsfig,endnotes}

\usepackage{silence}
\usepackage{comment}
\usepackage{multirow}
\usepackage[hyphens]{url}
\usepackage[hypertexnames=false,bookmarks=false,colorlinks]{hyperref}
\usepackage[hyphenbreaks]{breakurl}
\usepackage{xcolor}
\hypersetup{
  colorlinks,
  linkcolor={red!50!black},
  citecolor={blue!50!black},
  urlcolor={blue!50!black}
}
\usepackage[labelfont=bf,font=small,belowskip=-10pt,aboveskip=0pt]{caption}
\usepackage{amsmath,amsopn,amssymb}
\usepackage{endnotes,microtype,xspace,graphicx,fancyvrb,multirow}
\usepackage{booktabs}
\usepackage{array,underscore,relsize}
\usepackage[T1]{fontenc}
\usepackage{times}
\usepackage{fancyhdr,lastpage}
\usepackage{enumitem}
\usepackage{soul}
\usepackage{float}
\usepackage{balance}
\usepackage{textcomp}
\usepackage{subfig}
\usepackage{graphicx}
\usepackage[compact,small]{titlesec}
\usepackage{titling}

\titlespacing*{\section}{0pt}{0.3\baselineskip}{0.3\baselineskip}
\titlespacing*{\subsection}{0pt}{0.2\baselineskip}{0.2\baselineskip}

\usepackage{fp}
\usepackage{siunitx}

\usepackage{xcolor}

\begin{document}

\date{}

\title{Correct, Fast Remote Persistence}

\definecolor{mybrown}{rgb}{0.50, 0.02, 0.03}
\newcommand*\samethanks[1][\value{footnote}]{\color{mybrown}\footnotemark[#1]}

\author{
 {\rm Sanidhya Kashyap\thanks{Work done when author was at Oracle Labs.}}\\
 Georgia Institute of Technology
\and
 {\rm Dai Qin\samethanks}\\
 University of Toronto
\and
 {\rm Steve Byan \hspace{0.1in} Virendra J. Marathe}\\
 Oracle Labs
\and {\rm Sanketh Nalli}\\
  Oracle
} 

\maketitle

\thispagestyle{empty}

\def\Snospace~{\S{}}
\renewcommand*\sectionautorefname{\Snospace}
\def\sectionautorefname{\Snospace}
\def\subsectionautorefname{\Snospace}
\def\subsubsectionautorefname{\Snospace}
\def\chapterautorefname{\Snospace}
\newcommand{\subfigureautorefname}{\figureautorefname}

\newcommand{\yes}{Y}
\newcommand{\no}{}

\newcommand{\shl}{\ \cc{<}\cc{<}\ }
\newcommand{\shr}{\ \cc{>}\cc{>}\ }
\newcommand{\x}{$\times$\xspace}

\newcommand{\cc}[1]{\mbox{\textsc{#1}}}

\newcommand{\pmem}{\mbox{\cc{PM}}\xspace}
\newcommand{\rdma}{\mbox{\cc{RDMA}}\xspace}
\newcommand{\flush}{\mbox{\cc{Flush}}\xspace}

\newcommand{\qp}{\mbox{\cc{QPair}}\xspace}
\newcommand{\ddio}{\mbox{\cc{DDIO}}\xspace}
\newcommand{\nddio}{\mbox{\cc{$\neg$DDIO}}\xspace}

\newcommand{\Atomic}{\mbox{\cc{Atomic}}\xspace}
\newcommand{\cas}{\mbox{\cc{CAS}}\xspace}
\newcommand{\fna}{\mbox{\cc{FAA}}\xspace}
\newcommand{\Write}{\mbox{\cc{Write}}\xspace}
\newcommand{\WriteAtomic}{\mbox{\cc{Write$_{atomic}$}}\xspace}
\newcommand{\Read}{\mbox{\cc{Read}}\xspace}
\newcommand{\Send}{\mbox{\cc{Send}}\xspace}
\newcommand{\WriteID}{\mbox{\cc{WriteImm}}\xspace}

\newcommand{\RQWRB}{\mbox{\cc{RQWRB}}\xspace}

\newcommand{\ra}[1]{\renewcommand{\arraystretch}{#1}}

\newcommand{\rlog}{\mbox{\cc{RemoteLog}}\xspace}

\section*{Abstract}

Persistence of updates to remote byte-addressable persistent memory
(\pmem), using \rdma\ operations (\rdma\ updates), is a poorly
understood subject.  Visibility of \rdma\ updates on the remote server
is not the same as persistence of those updates.  The remote server's
configuration has significant implications on what it means for
\rdma\ updates to be persistent on the remote server.  This leads to
significant implications on methods needed to correctly persist those
updates.  This paper presents a comprehensive taxonomy of system
configurations and the corresponding methods to ensure correct remote
persistence of \rdma\ updates.  We show that the methods for correct,
fast remote persistence vary dramatically, with corresponding
performance trade offs, between different remote server
configurations.

\section{Introduction}

Byte addressable persistent memory (\pmem) such as Intel and Micron's
3D XPoint\texttrademark{}~\cite{3dxpoint} promises to make local
persistence faster than the state-of-the-art NAND flash by orders of
magnitude.  Furthermore, \pmem{}'s \emph{byte addressability} promises
to fundamentally change the way applications represent and manage
persistent data.  At the same time, modern \emph{Remote Direct Memory
Access (\rdma)} network fabrics (InfiniBand~\cite{ibta,ib-spec},
RoCE~\cite{ib-rocev2-annex}, and
iWARP~\cite{bestler-rfc5043,culley-rfc5044,pinkerton-rfc5042,rdma,shah-rfc5041})
are bringing network access latencies down to singleton microseconds.
They offer similar byte addressable remote memory access.  The
confluence of these two technologies in enterprise and distributed
systems, where high availability is critically important, is
inevitable.

Some early work~\cite{lu17,zhang15} has indeed demonstrated the
synergistic performance benefits \rdma\ and \pmem\ can deliver to
distributed, highly available applications.  However, little is
understood on how persistence of \rdma\ updates to remote \pmem\ can
be correctly achieved.  For instance, mere receipt of a completion
notification of a \rdma\ \Write\ to remote \pmem\ does not necessarily
mean that the \Write\ has persisted on the remote \pmem.  In fact,
persistence of remote
\pmem\ updates, using \rdma\ operations, truly depends on the
configuration of the remote server.  Absent the recipe to correctly
persist a remote update, serious consistency problems can emerge in
distributed applications in the face of failures.  

To the best of our knowledge, there is no comprehensive analysis on
system configurations and their implications on methods to correctly,
and efficiently, persist updates using
\rdma\ operations.  Douglas~\cite{douglas15} provides an
enumeration of some remote server configurations and corresponding
remote persistence recipes.  However, it lacks comprehensiveness.
Furthermore, certain system configurations assumed in Douglas'
categorization are not even relevant to today's state-of-the-art
system support for \pmem~\cite{rudoff16}.  Recent effort by the
InfiniBand Trade Association (IBTA) standards community proposes
extensions to \rdma\ operations to enable remote
persistence~\cite{burstein18,grun18}.  However, we show that the
proposed extensions cover correct, fast remote persistence for a
somewhat narrow set of remote server configurations.

This paper presents a comprehensive taxonomy of remote server
configurations that has significant implications on the methods
required to correctly persist \rdma\ updates to the remote
server's \pmem.  Our taxonomy breaks down configurations along three
axes: (i) the notion of a \emph{persistence domain} in a system -- the
portion of the memory hierarchy and RDMA-capable Network Interface
Card (RNIC) buffers that are effectively persistent; (ii) enablement of
optimizations to direct \rdma\ updates to the remote server's
processor cache~\cite{arm-cache-stashing,arm-dynamiq}, also called
\emph{Data Direct I/O (\ddio)} by Intel~\cite{ddio}; and (iii) location of
receive queue work request buffers (\RQWRB{}s) of \rdma\ connection
endpoints, called \emph{Queue Pairs (\qp{}s)}, on the remote server --
either in the server's DRAM or its \pmem.  

We incorporate not only existing \rdma\ operations in our analysis
(\rdma\ \Write, \rdma\ \WriteID, and \rdma\ \Send), but also
operations and extensions newly proposed by the IBTA standards
community~\cite{burstein18,grun18} -- \rdma\ \flush\ and
\emph{non-posted} \rdma\ \Write.

Our thorough analysis of the server configuration space and
\rdma\ operations has led us to $10$ distinct methods for remote
persistence of singleton \rdma\ updates.  Our analysis leads us to
some interesting, and surprising, methods of remote persistence that
let clients treat the traditionally two-sided \rdma\ \Send\ operation
as a one-sided \rdma\ operation.

We also examine correct persistence of \emph{compound} updates.  We
try to address the following question: What is the way to correctly,
and efficiently, persist causally dependent updates on the remote
server's \pmem?  (Much of the IBTA standards community discussion has
revolved around this question~\cite{burstein18,grun18}.)  We find $9$
additional methods for correct and efficient remote persistence for
compound updates.  Through our analysis we precisely pin-point the
correct and efficient method of remote persistence for each of the
combined $72$ different scenarios considered in our work.  The
programmer must carefully apply the correct remote persistence method
for a given remote server configuration.  Application of an incorrect
persistence method may lead to worse performance, or even critical
data inconsistencies in the face of failures.

We evaluate our methods of remote persistence using a ubiquitous
workload that manifests in most distributed and highly available
systems -- \emph{log replication (\rlog)}.  \rlog\ serves as the test
bed for both singleton and compound \rdma\ update persistence.  Our
evaluation shows significant performance trade offs between the
various methods of remote persistence, with a general indication that,
for \rlog, remote persistence via one-sided \rdma\ operations performs
significantly better than a message passing based approach using
\rdma\ \Send{}s.

We first present some background in \autoref{sec:background} that
describes pertinent system support for \pmem, the high level
architecture of a system that participates in a \rdma\ network,
various pertinent \rdma\ operations (current, and new ones proposed by
the IBTA standards community~\cite{burstein18,grun18}), and recent
technological advancements that have significant implications on
remote persistence.  In~\autoref{sec:taxonomy}, we describe our remote
persistence taxonomy and its implications on persisting RDMA updates
to \pmem.  We present a detailed qualitative analysis of the steps
needed for remote persistence.  Our taxonomy's methods of remote
persistence are based in part on recent \rdma\ extensions proposed by
the IBTA standards community~\cite{burstein18,grun18}.  We treat
persistence of \emph{singleton} and \emph{compound} remote updates
separately.
Our preliminary evaluation in~\autoref{sec:evaluation} indicates
significant performance trade offs between the various methods of
remote persistence.

\section{Background}
\label{sec:background}

\begin{figure}[t]
\includegraphics[width=\columnwidth]{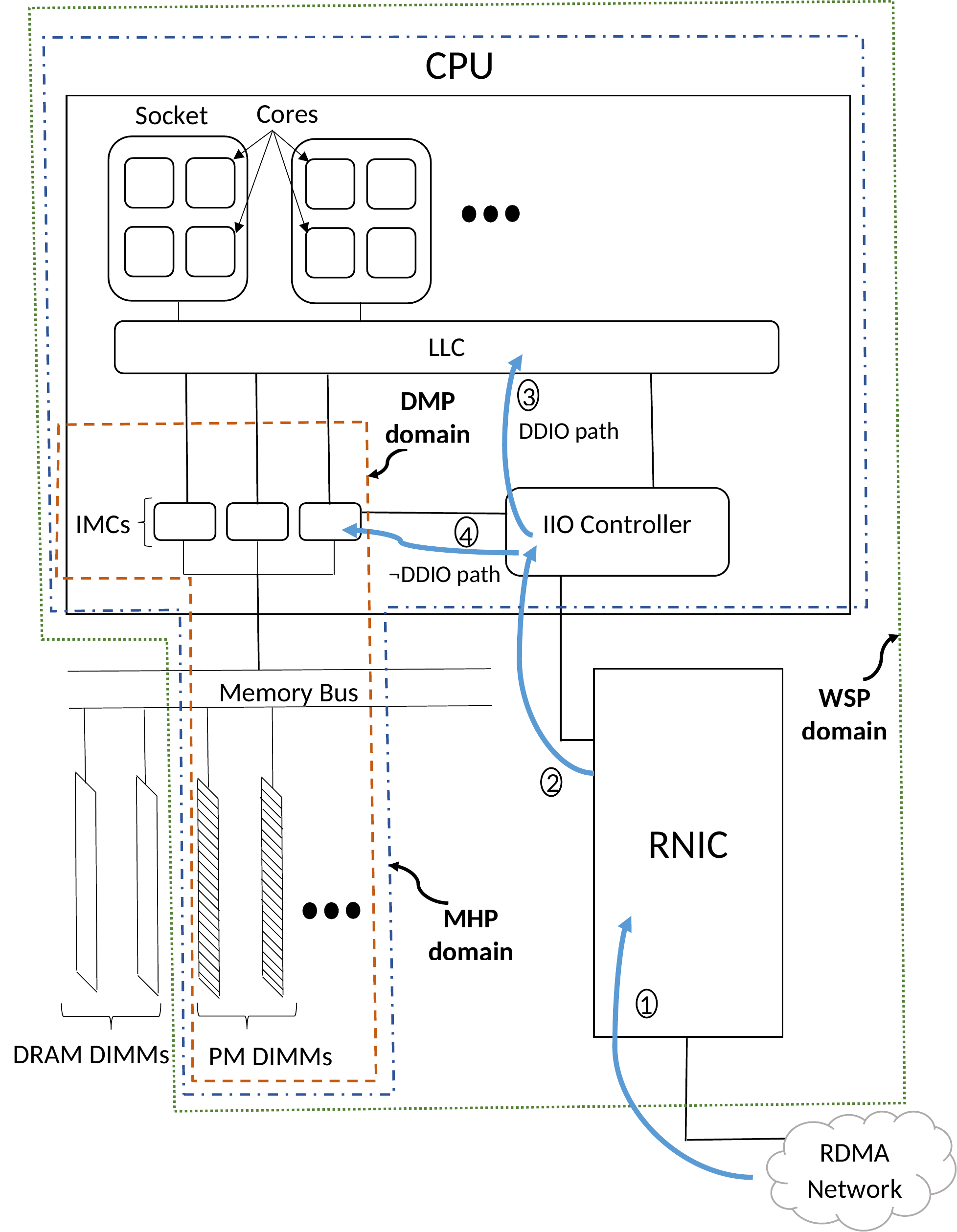}
\centering
\caption{Block Diagram of a computer connected to an RDMA network
  fabric.}
\label{fig:block-diagram}
\end{figure}

\paragraph{System Support for \pmem{}:}
\pmem\ supported systems are expected to become available
imminently~\cite{optane-dc}.  In these systems, \pmem\ will be
available in the DIMM form factor, alongside traditional DRAM DIMMs as
shown in \autoref{fig:block-diagram}.  Applications will be able to
perform \textsf{load}, \textsf{store}, and other memory access
instructions on \pmem\ that are typical of DRAM.  Furthermore,
processor vendors such as Intel~\cite{intel-isa} and
ARM~\cite{armv8-a} are providing extensions to enable ordering of
persistence of \textsf{store}s to \pmem\ -- e.g.~Intel's
\textsf{clflush-opt} and \textsf{clwb} instructions, and extensions to
existing instructions with fence semantics to enforce completion of
prior cache line flushes and writebacks (ARM has similar
extensions~\cite{armv8-a}).  Cache line flushes, writebacks and memory
fences guarantee that updated cache lines are evicted or copied to at
least the Integrated Memory Controller (IMC) buffers.  IMC buffer
entries are drained out to the \pmem\ DIMMs as scheduled by the 
memory controller. During a power failure, 
hardware features such as
\emph{Asynchronous DRAM Refresh (ADR)}~\cite{agiga-tech14,rudoff16}
are used to drain the IMC buffers to \pmem\ DIMMs.  \pmem\ is already
available in the industry in the form of NVDIMM-N type
DIMMs~\cite{nvdimm-n} -- DRAM backed by NAND flash using ultra
capacitors~\cite{agiga,micron,viking}.

\graphicspath{{./figures}}

\paragraph{Networked Computer Architectures:}
RNICs are becoming increasingly common in modern data centers.
\autoref{fig:block-diagram} depicts the high level architecture of
today's data center computers connected to each other via
\rdma\ network fabrics.  A computer communicates with another via the
RNIC.  The RNIC itself has internal buffers that can host data recently
sent or accessed by a remote node.  These buffers however are not coherent
with the host processor's cache~\cite{snia16}.

We now discuss the data flow between different parts of the system
caused by \rdma\ accesses (shown in~\autoref{fig:block-diagram}).  We
focus on \rdma\ \Write{}s here, but the same details apply to
\rdma\ \Send{}s, \Read{}s, and \Atomic{}s.  An incoming
\rdma\ \Write\ first lands in the RNIC buffers (step $1$
in~\autoref{fig:block-diagram}).  Thereafter, the RNIC posts (via DMA)
the \Write\ to the server's memory.  The data itself travels through
the processor's PCIe I/O controller.  The I/O controller was
traditionally a separate off-die chip between peripheral I/O devices
and the processor.  However, in recent years, processors have the
\emph{Integrated I/O (IIO)} feature that hosts the I/O controller on
the processor's die for a much more efficient data path between
peripheral I/O devices (e.g.~RNICs) and the processor.  IIO is
prevalent in modern processors, and has its private independent buffers.

The \rdma\ \Write{}'s data in practice moves from the RNIC buffers to
the IIO controller buffers (step $2$), after which it is moved to the
target memory DIMMs via the pertinent memory controllers (step $4$).
The received data can then be accessed by the server's
processor.  This necessarily leads to a high latency memory bus
transaction in the server to fetch the recently received data.
Recognizing these overheads, Intel introduced the notion of \emph{Data
  Direct I/O (\ddio)} in its processors that provides a data path from
the IIO directly to the server processor's L3 cache, rather than placing
the data in IMC buffers~\cite{ddio} (step
$3$).  A similar feature exists in processors by other vendors such as
ARM~\cite{arm-cache-stashing,arm-dynamiq}.

With \ddio, all incoming \rdma\ \Write{}s are likely to directly
reach the server processor's L3 cache; in high traffic scenarios,
writes may partially be written directly into the server's memory
DIMMs.  \ddio\ is a standard feature in modern Intel processors, and
can be switched off to avoid cache pollution if the application hosted
on the server is not expected to access recently received data through
its peripheral I/O devices.  As we will show
in~\autoref{sec:taxonomy}, \ddio\ plays an important role in remote
persistence.

\paragraph{RDMA Operations:}
The \rdma\ programming model~\cite{mellanox-rdma-manual,rdma,rdma-ext}
contains a \qp\ abstraction that is used to represent an endpoint of a
communication channel (connection) between computers (nodes) over and
RDMA network.  A connection management system establishes a connection
between \qp{}s of communicating nodes.  This includes designating the
type of the connection: Unreliable Datagram (UD), Unreliable
Connection (UC), or Reliable Connection (RC).  Since reliability
(messages are not lost) is critical to remote persistence, we assume
reliable connections in the rest of the paper unless mentioned otherwise.  All
\rdma\ data operations (detailed below) can be used in reliable connections.
In the rest of this paper, we call the remote server that receives
RDMA requests the \emph{responder}, and the requesting server is
called the \emph{requester}.

\rdma\ operations contain a ``two-sided'' message passing operation --
the \emph{\rdma\ \Send}.  The responder must receive and process the
sent message and potentially provide a response to the requester via
another \rdma\ \Send\ message directed to the requester.  The
responder's processor performs these tasks, which may not be desirable
for some applications~\cite{dragojevic14}.  The ``one-sided'' RDMA
operations -- \emph{\rdma\ \Read} and \emph{\rdma\ \Write} -- do not
require any participation from the responder's CPU; these operations
read and write remote memory respectively (the responder's RNIC uses
DMA to perform the reads and writes).  A hybrid
\emph{\rdma\ \Write\ with Immediate Data (\WriteID)} operation
performs a \Write\ on the responder's memory and delivers a message to
the responder with a 32-bit sized payload -- the \emph{immediate
  data}.  \rdma\ \WriteID\ does require the responder's CPU to process
the message with immediate data, like \rdma\ \Send.  \rdma\ also
contains one-sided \Atomic\ operations (\emph{\rdma\ Fetch-And-Add
  (\fna)} and \emph{\rdma\ Compare-And-Swap (\cas)}) -- 64-bit wide
atomic read-modify-write operations that can be used for
synchronization between remote requesters~\cite{rdma-ext,wei15}, but
can incur significant performance overheads~\cite{kalia16}.

All \rdma\ operations described above are asynchronous.  The requester
posts a \rdma\ \emph{work request} at a \qp\ and, 
if the request was flagged to return a completion notification, can
subsequently wait on a \emph{completion notification} for the request.  
The 
notification is generated at the requester once the operation
completes.  All operations that semantically generate a response
(\rdma\ \Read, \fna, and \cas) necessarily generate a completion
notification.  Work requests for other operations (\rdma\ \Send,
\Write, and \WriteID) can optionally be flagged for completion
notification. An application can use the RDMA ordering rules to infer
completion of operations that are not flagged for completion notification.

More recently~\cite{burstein18,grun18}, the IBTA standards community
decided to add a new \rdma\ operation -- the \emph{\rdma\ \flush}.
This is because simply receiving a completion notification, at the
requester, for a \rdma\ \Write{}/\Send{}/\WriteID\ does not guarantee
that the operation has become \emph{visible} (available in the memory
hierarchy) at the responder's end -- the operation may still reside in
the responder's RNIC buffers, or in the case of
iWARP~\cite{bestler-rfc5043,culley-rfc5044,pinkerton-rfc5042,rdma,shah-rfc5041},
may not even have left the requester.  As we will see
in~\autoref{sec:taxonomy}, a failure at the responder at such a
juncture could lead to loss of data residing in the responder's RNIC
buffers in some system configurations.  The new \rdma\ \flush\ operation
guarantees that all prior \rdma\ update operations, issued on the same
connection, become visible to the responder before the \flush{}'s
completion notification is received at the requester.

\paragraph{RDMA Operation Ordering:}
The \rdma\ standard specifies interesting ordering rules for
\rdma\ operations~\cite{ib-spec,rdma,rdma-ext}.  In particular, it
separates \rdma\ operations into two categories: (i) operations that
produce a return value that is consumed by the requester
(\rdma\ \Read, \cas, \fna, and now, \flush), informally referred to as
``non-posted'' operations in the community~\cite{burstein18,grun18};
and (ii) operations that do not require a return value (\rdma\ \Send,
\Write, and \WriteID), also referred to as ``posted'' operations.

The effects of non-posted operations are totally ordered with all prior
operations at the responder -- their effects are made visible at the
responder in the order they were issued at the requester.  Effects of posted
operations are totally ordered with each other.  However,
posted operations can be ordered at the responder \emph{before} prior
non-posted operations issued by the requester.

Visibility of updates to \pmem\ on a local system does not imply
persistence of those updates~\cite{kolli17,pelley14,snia13}.
Similarly, visibility of \rdma\ updates does not imply persistence of
those updates.  As a result, though \rdma\ updates become visible
in-order on a reliable connection, they may become persitent out-of-order.
\rdma\ \flush\ was introduced by IBTA precisely to enforce order of
persistence.  However, since \rdma\ \flush\ is a non-posted operation,
a subsequent posted \rdma\ operation can be ordered \emph{before} the
\flush\ leading to the out-of-order persistence problem.  To address
this problem, IBTA decided to add an \Atomic\ version of
\rdma\ \Write~\cite{burstein18,grun18} that guarantees atomicity
semantics similar to the other \Atomic\ operations (\fna\ and \cas).
In effect, an \Atomic\ \Write\ (\WriteAtomic) acts as a non-posted
operation -- it is ordered, at the responder, \emph{after} all
preceding posted and non-posted \rdma\ operations issued on the same
connection.  Furthermore, \WriteAtomic\ can be used to update up to
8-bytes atomically (all or nothing semantics).

Ordering enforcement between posted and non-posted \rdma\ operations
is not a new problem.  The pre-existing solution in the
\rdma\ standard is to tag posted \rdma\ operations with the
\emph{fence} flag.  This flag ensures that the fenced operation blocks
at the requester's end until all prior non-posted operations on the
same \qp\ complete -- their result is received at the requester's end.
Such fenced posted operations can be used in conjunction with
\rdma\ \flush\ to enforce ordering of persistence of two consecutive
\rdma\ updates.  However, it incurs the overhead of an extra
round-trip for the \rdma\ \flush, that the second \rdma\ update must
wait for.  \WriteAtomic\ eliminates that extra round-trip, and enables
pipelining of the dependent updates (separated by an intervening
\rdma\ \flush).

\section{Taxonomy of Remote Persistence}
\label{sec:taxonomy}

\subsection{Remote Server Configuration}
\label{sec:remote-srvr-config}

Persistence on the responder's end of RDMA based remote updates
depends on several characteristics of the responder's configuration.
We first describe these characteristics that enable us to build an
elaborate taxonomy of server configurations that have significant
implications on methods to perform remote update persistence on the
responder.  We break down the responder's configuration along three
axes: (i) the notion of a \emph{persistence domain} in a
system~\cite{marathe18,rudoff17}; (ii) enablement of optimizations to
direct \rdma\ updates to the remote server's processor cache
(e.g. \ddio\ in Intel's processors); and (iii) location of work
request buffers in the receive queue of \rdma\ \qp{}s on the responder
-- either on its DRAM or its \pmem.

\subsubsection{Persistence Domains}

Prior work~\cite{marathe18,rudoff17} has defined a persistence domain
as the portion of the memory hierarchy that is effectively persistent
-- writes reaching this part of the memory hierarchy are guaranteed to
persist across power failure and restart cycles.  We extend that
definition to include the RNIC's buffers as well.  We observe three
distinct persistence domains based on characteristics of a system's
configuration.

\begin{itemize}

\item[1.] \emph{DIMM and Memory controller Persistence (DMP):} This
  persistence domain includes the \pmem\ DIMMs as well as the
  integrated memory controller (IMC) buffers.  The IMC buffers are
  included in the persistence domain through hardware features
  such as Asynchronous DRAM Refresh (ADR)~\cite{rudoff16,rudoff17}
  that flush the IMC buffers to the \pmem\ DIMMs during a power
  failure event.  This is expected to be the dominant type of persistence
  domain in at least the near term.

\item[2.] \emph{Memory Hierarchy Persistence (MHP):} This persistence
  domain includes the entire memory hierarchy of the
  system~\cite{izraelevitz16,nawab15,rudoff17} -- all processor
  caches, store buffers, etc.  Here the assumption is that the system
  has enough residual power to flush processor buffers and caches to
  the \pmem\ DIMMs on a power failure event.  Custom system hardware
  and power supply equipment may be required to make this
  possible~\cite{rudoff17}.  MHP has significant implications on the
  programming model for \pmem\ since now the visibility of \textsf{store}
  instructions implies persistence -- there is no need for explicit
  cache line flushes and writebacks and corresponding persistence
  barriers~\cite{kolli17,pelley14}.  The RNIC buffers are however not
  included in the persistence domain.  As a result, there is still
  a need to perform \rdma\ \flush{}es.

\item[3.] \emph{Whole System Persistence (WSP):} This persistence
  domain encompasses the entire system, including the RNIC buffers.
  Battery backed systems can serve WSP~\cite{narayanan12}.  On a power
  failure, the caches of the processor as well as the buffers of all 
  peripheral devices will be flushed to corresponding regions in the
  \pmem\ DIMMs.  Since RNIC buffers become effectively persistent, WSP
  has interesting implications on the programming model for remote
  persistence in that it eliminates the need for explicit ordering barriers
  between posted \rdma\ operations. In WSP, the normal RDMA ordering semantics 
  now apply to persistence, once the data has been received by the responder
  RNIC.

\end{itemize}

\autoref{fig:block-diagram} depicts DMP, MHP, and WSP.  We do not
include a persistence domain limited to just the
\pmem\ DIMMs~\cite{douglas15,marathe18} since such persistence domains
are unlikely to be supported in real systems~\cite{rudoff16}.

\subsubsection{Implications of \ddio{} / Cache Stashing}

\ddio{} \cite{ddio} is the feature on Intel processors that enables
delivery of incoming \rdma\ data from the responder's RNIC
directly into the L$3$ cache.  A similar feature called \emph{cache
  stashing} is provided by ARM
processors~\cite{arm-cache-stashing,arm-dynamiq}.  While it does
provide the processor much more efficient access to incoming data,
for a system that supports just DMP, DDIO ends up keeping the inbound
data outside the persistence domain, that is, in the processor cache.    
Extra work needs to be done by the responder's processor to 
flush/writeback the arrived data to the persistence domain.  
Alternately, if \ddio\ is turned off at the responder, the inbound 
data will move to the memory controller, which is within the DMP 
domain.  If the responder supports MHP, arrival of the inbound data in
the responder's L$3$ cache implicitly persists the data.  For WSP,
arrival of inbound data in the RNIC buffers itself ensures
persistence.

\subsubsection{RDMA Queues and Work Requests}

Each \qp\ used for a \rdma\ connection internally contains two queues
-- the \emph{send queue} and the \emph{receive queue}.  Each of these
is a linked list of \emph{work requests (WRs)} created by the
enclosing application.  Each work request itself points to a
\emph{work request buffer (WRBs)} created by the application.  The
work request buffer contains application specific data.  For the send
queue, one or more work request buffers can be associated with a work
request.  The more pertinent queue for persistence of \rdma\ updates
is the receive queue.  The responder must preallocate work requests
(and their buffers) in a \qp{}'s receive queue in order to receive
work requests from the requester.  These work request buffers are
completely under the control of the application, which can allocated
them from either DRAM or \pmem.  As we shall see later, these
allocation choices have significant implications on remote persistence
of \rdma\ \Send\ requests to the extent that they can be treated by
the requester like one-sided \rdma\ updates.

We note that a \qp\ also is associated with a \emph{Completion Queue (CQ)} that
contains requester-side, RNIC generated, completion notifications for
operations explicitly marked to generate a completion notification.
For a posted operation, a completion notification is generated by the
requester's RNIC immediately after the responder's RNIC receives the
operation -- this is triggered from the lower level \rdma\ transport
layer, when the requester's RNIC receives an acknowledgment from the
responder's RNIC for the receipt of the operation at the responder's
end.  As noted earlier, for a non-posted operation, a completion
notification is generated at the requester only after the return value
of the non-posted operation is received from the responder.

Receipt of inbound two-sided \rdma\ operations (\rdma\ \Send\ and
\WriteID) results in creation of a \emph{receive completion}
notification in the \qp{}'s CQ.  This receive completion is however
generated \emph{after} the corresponding receive queue work request,
and its corresponding work request buffer, is populated and made
visible by the RNIC.

\begin{table}[t]
  \begin{center}
    \scriptsize{
      \resizebox{\columnwidth}{!}{
      \begin{tabular}{p{0.8in}cp{2.4in}}
\toprule
\multicolumn{1}{c}{\textbf{Config}} &
& \multicolumn{1}{c}{\textbf{Explanation}} \\
\midrule
{\bf DMP + \ddio\ + DRAM-RQWRB} &$\rightarrow$& DMP, with
\ddio\ turned on, and RQWRB placed in DRAM.\\

{\bf DMP + \ddio\ + \pmem{}-RQWRB} &$\rightarrow$& DMP, with
\ddio\ turned on, and RQWRB placed in \pmem.\\

{\bf DMP + \nddio\ + DRAM-RQWRB} &$\rightarrow$& DMP, with
\ddio\ turned off, and RQWRB placed in DRAM.\\

{\bf DMP + \nddio\ + \pmem{}-RQWRB}&$\rightarrow$& DMP, with
\ddio\ turned off, and RQWRB placed in \pmem.\\

{\bf MHP + \ddio\ + DRAM-RQWRB} &$\rightarrow$& MHP, with
\ddio\ turned on, and RQWRB placed in DRAM.\\

{\bf MHP + \ddio\ + \pmem{}-RQWRB} &$\rightarrow$& MHP, with
\ddio\ turned on, and RQWRB placed in \pmem.\\

{\bf MHP + \nddio\ + DRAM-RQWRB} &$\rightarrow$& MHP, with
\ddio\ turned off, and RQWRB placed in DRAM.\\

{\bf MHP + \nddio\ + \pmem{}-RQWRB} &$\rightarrow$& MHP, with
\ddio\ turned off, and RQWRB placed in \pmem.\\

{\bf WSP + \ddio\ + DRAM-RQWRB} &$\rightarrow$& WSP, with
\ddio\ turned on, and RQWRB placed in DRAM.\\

{\bf WSP + \ddio\ + \pmem{}-RQWRB} &$\rightarrow$& WSP, with
\ddio\ turned on, and RQWRB placed in \pmem.\\

{\bf WSP + \nddio\ + DRAM-RQWRB} &$\rightarrow$& WSP, with
\ddio\ turned off, and RQWRB placed in DRAM.\\

{\bf WSP + \nddio\ + \pmem{}-RQWRB} &$\rightarrow$& WSP, with
\ddio\ turned off, and RQWRB placed in \pmem.\\
\bottomrule
\end{tabular}

      }
    }
  \end{center}
  \caption{Remote server configurations.
    \RQWRB is the Receive Queue Work Request Buffer.}
  \label{tab:configs}
\end{table}


\noindent Our analysis above leads us to twelve distinct remote server
configurations as shown in~\autoref{tab:configs}.


\subsection{Persisting Singleton \rdma\ Updates}
\label{sec:singleton-updates}

\begin{table*}[t]
  {\scriptsize
    \begin{center}

      \renewcommand{\tabcolsep}{1pt}
        \begin{tabular}{| l | l | l | l | l | l | l | l | l | l |}
    \hline
    & \multicolumn{3}{c|}{\bf DMP}
    & \multicolumn{3}{c|}{\bf MHP}
    & \multicolumn{3}{c|}{\bf WSP} \\
    \hline
    & & & & & & & & & \\
    & \multicolumn{1}{c|}{\bf Write}
    & \multicolumn{1}{c|}{\bf WriteImm}
    & \multicolumn{1}{c|}{\bf Send}
    & \multicolumn{1}{c|}{\bf Write}
    & \multicolumn{1}{c|}{\bf WriteImm}
    & \multicolumn{1}{c|}{\bf Send}
    & \multicolumn{1}{c|}{\bf Write}
    & \multicolumn{1}{c|}{\bf WriteImm}
    & \multicolumn{1}{c|}{\bf Send} \\
    & & & & & & & & & \\
    \hline
    \multirow{1}{*} {\bf {}\ddio\ $+$} &
      Rq Write(a) & Rq WriteImm(a) & Rq Send(a) & Rq Write(a) & Rq WriteImm(a) & Rq Send(a) & Rq Write(a) & Rq WriteImm(a) & Rq Send(a) \\
      {\bf DRAM-RQWRB} & Rq Send(\&a) & Rsp Receive(\&a) & Rsp Receive(a) & Rq Flush & Rq Flush & Rsp Receive(a) & Rq Comp$_{Write(a)}$ & Rq Comp$_{WriteImm(a)}$ & Rsp Receive(a) \\
      & Rsp Receive(\&a) & Rsp flush(\&a) & Rsp copy(a) + & Rq Comp$_{Flush}$ & Rq Comp$_{Flush}$ &  Rsp copy(a) & & & Rsp copy(a) \\
      & Rsp flush(\&a) & Rsp Send(ack) & \ \ \ \ \ \ \ \ flush(\&a) & & & Rsp Send(ack) & & & Rsp Send(ack) \\
      & Rsp Send(ack) & Rq Receive(ack) & Rsp Send(ack) & & & Rq Receive(ack) & & & Rq Receive(ack)  \\
      & Rq Receive(ack) & & Rq Receive(ack) & & & & & & \\

    \hline
    \multirow{1}{*} {\bf {}\ddio\ $+$} &
      As above & As above & As above & As above & As above & Rq Send(a) & As above & As above & Rq Send(a) \\
      {\bf \pmem-RQWRB} & & & & & & Rq Flush & & & Rq Comp$_{Send(a)}$ \\
      & & & & & & Rq Comp$_{Flush}$ & & & \\

    \hline
    \multirow{1}{*} {\bf {}\nddio\ $+$} &
      Rq Write(a) & Rq WriteImm(a) & As above & As above & As above & Rq Send(a) & As above & As above & Rq Send(a) \\
      {\bf DRAM-RQWRB} &  Rq Flush$^*$ & Rq Flush &  & & & Rsp Receive(a) & & & Rsp Receive(a) \\
      &  Rq Comp$_{Flush}$ & Rq Comp$_{Flush}$ & & & & Rsp copy(a)  & & & Rsp copy(a) \\
      & & & & & & Rsp Send(ack) & & & Rsp Send(a) \\
      & & & & & & Rq Receive(ack) & & & Rq Receive(a) \\

    \hline
    \multirow{1}{*} {\bf {}\nddio\ $+$} &
      As above & As above & Rq Send(a) & As above & As above & Rq Send(a) & As above & As above & Rq Send(a) \\
      {\bf \pmem-RQWRB} &  &  & Rq Flush & & & Rq Flush & & & Rq Comp$_{Send(a)}$ \\
      &  &  & Rq Comp$_{Flush}$ & & & Rq Comp$_{Flush}$ & & & \\
    \hline
  \end{tabular}


  \renewcommand{\tabcolsep}{6pt}

  \begin{tabular}{lll}
    RQWRB $=$ Receive Queue Work Request Buffer & Rq = Requester & Rsp = Rsponder\\
    copy $=$ local memcpy at the Responder & flush $=$ local cache line flush at the Responder\\
    Rq Completion = Receipt of completion notification at requester
    & Receive = Requester/Responder receives message\\
  \end{tabular}

      \scriptsize{$^*$On some systems, a message exchange, using \rdma\ \Send,
        can possibly be more efficient than a \rdma\ \flush.
      }

    \end{center}

\caption{Taxonomy for Singleton Updates (value $a$) using
  \rdma\ operations to location $\&a$ in the responder's \pmem.  Each
  row in the table corresponds to a remote server (responder)
  configuration with \ddio\ turned on or off and the RQWRB resident in
  DRAM or \pmem.  Each column represents the primary operation used to
  implement the remote update (\rdma\ \Write, \rdma\ \WriteID, or
  \rdma\ \Send), further grouped by the persistence domain configured
  on the responder -- DMP, MHP, WSP.  Value $a$ is written by the
  requester at the responder; $\&a$ represents the address of the
  target \pmem\ memory block at the responder.}

\label{tab:singleton-update}
}
\end{table*}

We first consider the case of singleton remote updates comprising
update of just one contiguous block of data in the responder's \pmem.
The block can range from $1 - 2^{31}$ bytes in size, the permissible
size for a \rdma\ \Send, \Write, and \WriteID\ on a reliable connection.
\autoref{tab:singleton-update} depicts the full taxonomy of remote
persistence of singleton remote updates for the twelve different
responder configurations enumerated in~\autoref{tab:configs}.  We
break down our discussion of remote persistence by the persistence
domain configured on the responder.

\paragraph{\bf DMP:}
The DMP domain comprises just the \pmem\ DIMMs and the host
processor's IMC buffers (see~\autoref{fig:block-diagram}).  As a
result, if \ddio\ is turned on at the responder, incoming
\rdma\ updates may very well go into its processor cache, which is not
a part of its DMP.  As a result, the requester must send a message to
the responder to flush those updates to DMP.  It is clear that remote
persistence is not reliably possible with just one-sided
\rdma\ operations (\rdma\ \Write\ and \rdma\ \flush).  The alternative
shown in~\autoref{tab:singleton-update} is for the requester to
perform an \rdma\ \Write\, followed by a \rdma\ \Send\ that informs
the responder that a \Write\ has just happened, and needs to be
flushed out of the responder's processor cache.  After performing the
flush, the responder sends back a response to the requester informing
completion of remote persistence.  Remote persistence is guaranteed to
have happened from the requester's perspective when it receives the
response.

With \rdma\ \Send\ we assume that the sent message contains the
target location to update in the responder's \pmem\ and the value that
needs to be written to that location. We thus use the
standard message passing idiom to enforce persistence of remote
updates.  It however leads to a copy of the payload on the
responder's side -- from the RQWRB of the responder to the
target memory location of the responder (done using local copy
followed by local flush shown in~\autoref{tab:singleton-update}).

\rdma\ \WriteID\ is much more elegant in that it eliminates the
copying overhead (we assume that the write in \WriteID\ is directed to
a location in \pmem).  However it does require the responder's
processor to perform local flushes of the updated cache lines.
Furthermore, \rdma\ \WriteID\ has the limitation that the target
address must be identifiable using just the 32-bit immediate data
embedded in the message delivered to the responder, which may not be
possible in some application contexts.

Remote persistence gets more interesting when \ddio\ is turned off.
First, remote persistence using one-sided operations -- \rdma\ \Write,
 \rdma\ \WriteID, and \rdma\ \flush\ -- becomes possible (we assume that
\rdma\ \Write\ or \rdma\ \WriteID\ updates a location in the responder's \pmem).  The
requester must wait for the completion notification for the
\rdma\ \flush, which is received only after its \flush\ has taken
effect on the responder.  Prior work~\cite{douglas15,snia16} has
described this particular case.  We note however that in some
\rdma\ network and RNIC implementations, an \rdma\ \flush\ could possibly
have higher latency than a two-sided message exchange using
\rdma\ \Send{}s.

With \ddio\ turned off, if the RQWRB resides in DRAM, the method of
remote persistence with \rdma\ \Send\ follows the
traditional messaging passing idiom.  However, if the RQWRB resides in
\pmem\ an interesting possibility emerges: Since the RQWRB resides in
\pmem, a message received in the responder's DMP domain is effectively
persistent.  This message can survive power failure cycles, and the
enclosing application's recovery subsystem can be designed to access
these messages and persist their effects.  As a result, the requester
needs to simply persist the data from \rdma\ \Send\ operations
on the responder, which is done by issuing a \rdma\ \flush.  In
applications where processing of these persistent messages is not
possible during recovery (because of additional missing context that
was hosted in DRAM), the standard two-sided message exchange idiom
would be the only viable alternative.

\paragraph{\bf MHP:}
MHP refers to persistence of the whole memory hierarchy.  Since the
RNIC buffers are not a part of the MHP domain, \rdma\ \flush\ is required for
remote persistence using just one-sided operations.  That is the case
not only with \rdma\ \Write, but also with \rdma\ \WriteID.  In the
latter case, we assume that application correctness is not compromised
even if the immediate data delivered with the \rdma\ \WriteID\ is lost due to
a power failure or application crash.  (If synchronous immediate data
delivery is required for application correctness, a different method
for remote persistence, most likely based on \rdma\ \Send, must be
used by the application.)  The requester can infer persistence of
remote updates when it receives a completion notification for the
\rdma\ \flush.

The \rdma\ \Send\ based method for persistence of remote updates
varies based on whether the RQWRB resides in DRAM or \pmem.  If in DRAM,
the method is the classic message passing based idiom.  Note however
that the responder's processor does not need to issue local cache line
flushes since completing local stores itself ensures persistence.  If
the RQWRB resides in \pmem, persisting just the \rdma\ \Send\ on the
responder may be sufficient for the requester to infer persistence of
the remote update.  Again, if the application requires a synchronous
handshake between the requester and responder, the classic message
passing idiom may be a better match, but may come with higher latency
for remote persistence.

\paragraph{\bf WSP:}
When even the RNIC buffers become effectively persistent, persistence
of remote updates may become quite simple.  
For the InfiniBand and RoCE RDMA transports, receipt of just the
completion notification of \rdma\ \Write\ and \rdma\ \WriteID\ at the
requester is sufficient to infer persistence of a remote update.  The
same method of remote persistence applies to the \rdma\ \Send\ based
approach if the responder's RQWRB resides in \pmem.  If not, the classic
message exchange idiom is required for remote persistence.

We note a key difference in the semantics of completion notifications
between
iWARP~\cite{bestler-rfc5043,culley-rfc5044,pinkerton-rfc5042,rdma,shah-rfc5041}
and InfiniBand/RoCE~\cite{ib-rocev2-annex,ib-spec,ibta}.  InfiniBand
and RoCE guarantee that the \rdma\ operation is received at least at
the responder's RNIC \emph{before} the corresponding completion
notification is generated at the requester.  iWARP, however, makes a
``weaker'' guarantee in that a completion notification is created as
soon as the operation reaches the requester's reliable transport layer
(TCP, or SCTP~\cite{bestler-rfc5043}).  As a result, the completion
notification may be received by the application on the requester's
side even before the operation is sent to the responder.  The
implication the iWARP semantics have on remote persistence for responders
supporting WSP is that \rdma\ \flush\ becomes necessary for correct
remote persistence.  The methods for remote persistence for WSP
essentially mimic the corresponding methods for remote persistence for
MHP on iWARP.

\subsection{Persisting Compound \rdma\ Updates}
\label{sec:compound-updates}

\begin{table*}[t]
{\scriptsize
\begin{center}

  \renewcommand{\tabcolsep}{1pt}
    \begin{tabular}{| l | l | l | l | l | l | l | l | l | l |}
    \hline
    & \multicolumn{3}{c|}{\bf DMP} 
    & \multicolumn{3}{c|}{\bf MHP} 
    & \multicolumn{3}{c|}{\bf WSP} \\
    \hline
    & & & & & & & & & \\
    & \multicolumn{1}{c|}{\bf Write} 
    & \multicolumn{1}{c|}{\bf WriteImm} 
    & \multicolumn{1}{c|}{\bf Send} 
    & \multicolumn{1}{c|}{\bf Write} 
    & \multicolumn{1}{c|}{\bf WriteImm} 
    & \multicolumn{1}{c|}{\bf Send} 
    & \multicolumn{1}{c|}{\bf Write} 
    & \multicolumn{1}{c|}{\bf WriteImm} 
    & \multicolumn{1}{c|}{\bf Send} \\
    & & & & & & & & & \\
    \hline
    \multirow{1}{*} {\bf {}\ddio\ $+$} & 
      Rq Write(a) & Rq WriteImm(a) & Rq Send(a,b) & Rq Write(a) & Rq WriteImm(a) & Rq Send(a,b) & Rq Write(a) & Rq WriteImm(a) & Rq Send(a,b) \\
      {\bf DRAM-RQWRB} &
      Rq Send(\&a) & Rsp Receive(\&a) & Rsp Receive(a,b) & Rq Write(b) & Rq WriteImm(b) & Rsp Receive(a,b) & Rq Write(b) & Rq WriteImm(b) & Rsp Receive(a,b) \\
      &
      Rsp Receive(\&a) & Rsp flush(\&a) & Rsp copy + & Rq Flush & Rq Flush &  Rsp copy(a,b) & Rq Comp$_{Write(b)}$ & Rq Comp$_{WriteImm(b)}$ & Rsp copy(a,b) \\
      &
      Rsp flush(\&a) & Rsp Send(ack) & \ \ \ \ \ \ \ flush(a,b) & Rq Comp$_{Flush}$ & Rq Comp$_{Flush}$ & Rsp Send(ack) & & & Rsp Send(ack) \\
      &
      Rsp Send(ack) & Rq Receive(ack) & Rsp Send(ack) & & & Rq Receive(ack) & & & Rq Receive(ack)  \\
      &
      Rq Receive(ack) & Rq WriteImm(b) & Rq Receive(ack) & & & & & & \\
      &
      Rq Write(b) & Rsp Receive(\&b) & & & & & & & \\
      &
      Rq Send(\&b) & Rsp flush(\&b) & & & & & & & \\
      &
      Rsp Receive(\&b) & Rsp Send(ack) & & & & & & & \\
      &
      Rsp flush(\&b) & Rq Receive(ack) & & & & & & & \\
      &
      Rq Send(ack) & & & & & & & & \\
      &
      Rq Receive(ack) & & & & & & & & \\

    \hline

    \multirow{1}{*} {\bf {}\ddio\ $+$} & 
      As above & As above & As above & As above & As above & Rq Send(a,b) & As above & As above & Rq Send(a,b) \\
      {\bf \pmem-RQWRB} & 
      & & & & & Rq Flush & & & Rq Comp$_{Send(a,b)}$ \\
      & 
      & & & & & Rq Comp$_{Flush}$ & & & \\
    
    \hline
    
    \multirow{1}{*} {\bf {}\nddio\ $+$} &
      Rq Write(a) & Rq WriteImm(a) & As above & As above & As above & Rq Send(a,b) & As above & As above & Rq Send(a,b)\\
      {\bf DRAM-RQWRB} &  
      Rq Flush$^*$ & Rq Flush & & & & Rsp Receive(a,b) & & & Rsp Receive(a,b) \\
      &  
      Rq Write$_{atomic}$(b) & Rq Comp$_{Flush}$ & & & & Rsp copy(a,b)  & & & Rsp copy(a,b) \\
      & 
      Rq Flush & Rq WriteImm(b) & & & & Rsp Send(ack) & & & Rsp Send(ack \\
      & 
      Rq Comp$_{Flush}$ & Rq Flush & & & & Rq Receive(ack) & & & Rq Receive(ack) \\
      & 
      & Rq Comp$_{Flush}$ & & & & & & &  \\

    \hline

    \multirow{1}{*} {\bf {}\nddio\ $+$} &
      As above & As above & Rq Send(a,b) & As above & As above & Rq Send(a,b) & As above & As above & Rq Send(a,b) \\
      {\bf \pmem-RQWRB} &  
      & & Rq Flush & & & Rq Flush & & & Rq Comp$_{Send(a,b)}$ \\
      &  
      & & Rq Comp$_{Flush}$ & & & Rq Comp$_{Flush}$ & & & \\
    \hline
  \end{tabular}

  \renewcommand{\tabcolsep}{6pt}

  \begin{tabular}{lll}
    RQWRB $=$ Receive Queue Buffer & Rq = Requester & Rsp = Responder\\

    \WriteAtomic\ $=$ \rdma\ Atomic \Write\ & copy $=$ local memcpy at
    the Responder & flush $=$ local cache line flush\\

    Rq Comp = Receipt of completion notification at requester &
    Receive = Requester/Responder receives message\\
  \end{tabular}


\scriptsize{$^*$On some systems, a message exchange, using \rdma\ \Send,
  can possibly be more efficient than a \rdma\ \flush.
}

\end{center}

\caption{Taxonomy for Compound Updates using \rdma\ operations.  The
  above taxonomy orders remote persistence of two updates -- $a$
  followed by $b$.}

\label{tab:compound-updates}
}
\end{table*}

Ordering of consecutive updates is foundational to achieve data
consistency.  \rdma\ based updates are no different.  We now focus on
the methods programmers can use to enforce correct order of
persistence of consecutive updates using \rdma\ operations.  We want
to ensure that if a requester is posting two \emph{strictly ordered}
updates, $a$ followed by $b$, to the responder's \pmem, those updates
are persisted at the responder in the same order.  \emph{Log append}
is a canonical example of such dependent updates -- the log record at
the remote log's tail must first be updated and persisted, before
advancing the log's tail pointer and persisting it.
\autoref{tab:compound-updates} shows the recipes to enforce the
correct order of persistence on the responder.

\paragraph{\bf DMP:}
With \ddio\ turned on, dependent \rdma\ \Write{}s must be separated by
a message exchange between the requester and responder.  The responder
must also flush the affected cache lines for the first
\rdma\ \Write\ to ensure its persistence before sending back an
acknowledgment to the requester.  Furthermore, the second
\rdma\ \Write\ must also be followed by another identical message
exchange to inform the requester about persistence of the second
\rdma\ \Write.  We observe similar set of operations required for
remote persistence of consecutive \rdma\ \WriteID{}s; the
\WriteID\ itself ends up performing a write followed by delivery of a
message to the RQWRB of the responder.

\rdma\ \Send\ is perhaps the more effective way of performing remote
updates since both updates ($a$ and $b$) can be packaged in a single
message.  The responder must first write and flush $a$ before writing
and flushing $b$.  A receipt of an acknowledgment from the responder
informs the requester that the two updates have persisted on the
responder's \pmem.  The \rdma\ \Send\ based approach does however lead
to two copies of the updates at the responder's end -- in the RQWRB
and the final target.  As a result, the above two approaches may be
more efficient for coarse grained updates.  The methods for remote
persistence of dependent updates remains the same as above even when
the RQWRB resides in \pmem\ since \ddio\ can push the remote updates
to the responder's processor cache, which the responder must flush locally
to its DMP domain.

We observe interesting implications if \ddio\ is turned off on the
responder.  When the \rdma\ \Write\ operation is used to perform the
remote updates, there are two possible means of ordering the
persistence of the two updates $a$ and $b$.
\autoref{tab:compound-updates} shows just one alternative that aligns
with the canonical log append example mentioned above -- the second
update is a fine-grain write, to the log's tail, that atomically
updates at most $8$ bytes.  The \rdma\ \flush\ after the first
\rdma\ \Write\ ensures that the write will be flushed to the
responder's \pmem.  The use of \rdma\ \WriteAtomic\ for the second
write ensures that, at the responder, it is ordered \emph{after} the preceding
\rdma\ \flush.  The second subsequent \rdma\ \flush\ makes sure that
the \WriteAtomic\ persists before the completion notification for the
second \flush\ is received at the requester.

If the second update is more than $8$ bytes long, a \rdma\  \WriteAtomic\ will
not work.  In that case, the
requester must wait for the completion notification of the first
\rdma\ \flush\ before before issuing the second \rdma\ \Write.  
As mentioned earlier, in some \rdma\ fabric and RNIC
implementations, a message exchange based notification of completion
of remote persistence (similar to the approach taken above in the case
where \ddio\ is turned on) may perform better than the use of
\rdma\ \flush.

With \ddio\ turned off, \rdma\ \WriteID\ can be used as somewhat of a
one-sided operation in that the responder does not need to send back
an acknowledgment message to the requester after either of the two
\rdma\ \WriteID{}s. 
(We again assume that application correctness is not compromised
even if the immediate data delivered with the \rdma\ \WriteID\ is lost due to
a power failure or application crash.)
However, since there is no \Atomic\ version of
\WriteID, the requester must wait for the completion notification of
its first \rdma\ \flush\ before performing the second (dependent)
\rdma\ \WriteID.  Thereafter the second \rdma\ \flush\ and its
corresponding completion notification informs the requester that the
second update has also remotely persisted.

For \rdma\ \Send\ based remote updates, the method described above for
the \ddio\ case will work correctly if the RQWRB resides in DRAM.
However, if the RQWRB resides in \pmem, the \rdma\ \Send\ can be used
like a one-sided operation.  This is because using
\rdma\ \flush\ ensures that the sent message resides in the
\pmem\ location of the corresponding RQWRB at the responder.  This
implicitly presists the compound update, which can survive an
immediate power failure at the responder.  The application's recovery
subsystem can be used to find and apply the sent message to the
correct locations in the responder's \pmem.  The requester can infer
that the compound update has persisted on receipt of completion
notification for the \rdma\ \flush.

\paragraph{\bf MHP:}
For MHP, visibility of \rdma\ updates at the responder is equivalent
to persistence, as long as the updates are directed to the remote
\pmem.  Existing \rdma\ ordering
semantics~\cite{ib-spec,rdma,rdma-ext} guarantee in-order visibility
of consecutive posted updates.  As a result, two dependent updates can
be pipelined back-to-back as \rdma\ \Write{}s.  However, since the
writes need to be flushed from the responder's RNIC buffers to its
memory hierarchy, a \rdma\ \flush\ is needed.  The requester can
conclude ordered remote persistence of the two \Write{}s upon receipt of
the completion notification for the \flush.  \rdma\ \WriteID\ can be
treated as a one-sided operation by the requester and used in a way
similar to \rdma\ \Write\ for MHP.

For \rdma\ \Send, the requester can send a compound message containing
both the dependent updates, which are applied (using local stores) by
the responder in the expected order.  No local cache line flushes are
required at the responder because of MHP.  However, the responder must
send an acknowledgment to the requester informing the latter of
persistence of the compound update.  If the RQWRB resides in \pmem,
\rdma\ \Send\ can be treated as a one-sided operation by the
requester.  As a result, a subsequent \rdma\ \flush\ followed by
receipt of completion of the \flush\ is all the requester needs to
infer that the compound update has persisted on the responder's end.
However, note that the requester cannot immediately try to read the
responder's affected memory without additional coordination with the
responder.  If no coordination takes place, the requester might end up
reading a stale value from the responder if the preceding
\rdma\ \Send\ was not applied at the responder.

\paragraph{\bf WSP:}
The \rdma\ reliable connection guarantees ordered delivery of update requests
(\rdma\ \Write, \WriteID, and \Send) at the responder's RNIC.  As a
result, with \rdma\ \Write\ based updates, the requester can simply
post the \Write\ requests in the expected order.  The requester can
assume remote persistence of the two updates on receipt of a
completion notification of the second \rdma\ \Write.
\rdma\ \WriteID\ can be treated like a one-sided operation for WSP,
allowing simple back-to-back issuance of \WriteID{}s for ordered
remote updates.  Again the requester simply needs to wait for the
completion notification for the second \WriteID\ to infer remote
persistence.  Similar operation sequence can be used with
\rdma\ \Send\ if the RQWRB resides in \pmem.  However, both the
dependent updates can be packaged in a single \Send\ message.  (As
noted earlier in the singleton update case, \rdma\ \flush\ will be
required for the iWARP transport protocol.)  If the responder's RQWRB
resides in DRAM, the typical message passing idiom is needed to ensure
remote persistence of the two updates.

\subsection{Discussion}
\label{sec:discussion}

Our analysis above is intended to provide guidance to application
developers for correct remote persistence.  It is clear that the
method for remote persistence using \rdma\ operations varies
significantly between the twelve different configurations detailed
above.  We make a few interesting observations based on our analysis
of remote persistence methods for these various remote server
configurations.

First, the \ddio\ feature was originally introduced to improve
performance of applications that used the \rdma\ networking fabric.
We however find that the \ddio\ optimization gets in the way of
performing remote persistence using just the one-sided operations --
\rdma\ \Write\ and \rdma\ \flush\ -- in remote servers configured with
DMP, which will likely be a substantial portion, perhaps a majority,
of systems supporting \pmem\ in the near future.  Second, for
MHP and WSP, placing the RQWRB in \pmem, enables treatment of
\rdma\ \Send\ messages as one-sided operations, which
will likely lead to lower latency communication between the requester
and responder using these operations.  Third, with WSP, the new
\rdma\ \flush\ operation that is being discussed in the IBTA standards
community~\cite{burstein18,grun18} becomes unnecessary for
InfiniBand and RoCE network fabrics, although it is still required for iWARP.
Fourth, for compound
\rdma\ updates, we find that the new \WriteAtomic\ operation applies
to a narrow set of configurations in the whole taxonomy.
Lastly, while we described different methods of remote persistence for
different system configurations, methods such as \rdma\ \Send\ based
message passing for remote persistence are \emph{universal} in that
they can be used in \emph{all} system configurations.  However, as we
will see in~\autoref{sec:evaluation}, they come with a performance
penalty compared to remote persistence with just one-sided
\rdma\ operations.

We note that \rdma\ \flush\ and non-posted \rdma\ \Write\ are not
supported in today's RDMA protocol.  However, \rdma\ \flush\ can be
correctly emulated using \rdma\ \Read\ \cite{douglas15}.  This is
because \rdma\ \Read\ flushes the responder's RNIC's buffers the IIO
per the RDMA ordering rules
and then triggers a PCIe \Read\ at the responder's RNIC,
which in turn flushes the IIO buffers to memory~\cite{pcie3}.  
Non-posted \rdma\ \Write\ cannot be
correctly emulated by any existing \rdma\ operations at present.
Fenced \rdma\ \Write\ can be used for similar ordering behavior,
however, it adds an extra round-trip between the requester and
responder before the fenced \Write\ can be sent by the requester.

\emph{Torn writes} are always a data consistency concern for
persistence.  The concern is no different in \rdma\ based remote
persistence.  The application must ensure robustness against torn
writes via algorithmic techniques such as \emph{checksums} and
strictly ordered writes~\cite{burstein18,marathe18,volos11}, all of
which are very well understood in the literature.

\section{Evaluation}
\label{sec:evaluation}

Our taxonomy clearly demonstrates that the method to correctly ensure
persistence of \rdma\ updates varys signficantly between the various
system configurations.  We however would like to understand the
performance trade offs between these methods.  There are several key
questions we want to answer using our evaluation.  (i) Do the
different methods for remote persistence perform differently?  (ii) Is
there a significant enough performance gap between the universal
message passing based remote persistence and one-sided remote
persistence?  (iii) How much performance impact do the various
persistence domains have on remote persistence?  (iv) Does
\ddio\ affect remote persistence performance?  (v) Does placement of
\RQWRB\ in DRAM or \pmem\ matter to performance?

To answer these questions, we use a workload that is ubiquitous to
distributed systems that perform replication for high availability --
\emph{log replication}.  Log replication is arguably the dominant
method used to perform replication of updates to remote nodes in a
distributed system.

\subsection{Log Replication with \rlog}

Our benchmark, called \rlog, sets up a contiguous log at the server
end that is accessible to the client over a \rdma\ connection.
\rlog{}'s client repeatedly appends $10$ million log records to the
log.  Each append is made the \rdma\ based remote update and
persistence methods discussed in this paper.  We perform the
experiment for all the $72$ configurations from
\autoref{tab:singleton-update} and~\autoref{tab:compound-updates}.  We
report average log append latency at the end of each experiment.

\rlog{}'s append operation provides a test bed for both singleton
\rdma\ updates and compound \rdma\ updates.  Log appends happen at the
tail of the log.  This can be done using singleton \rdma\ updates by
encoding the log record with a checksum.  This checksum is used to
detect the tail of the log at the server -- the server detects the log
tail when its checksum fails.  Checksums are also an effective way to
detect data corruption.  Thus checksummed log records can enable a way
to do log appends using singleton \rdma\ updates.

Another dominant means of maintaining a log is by explicitly managing
the server's log tail pointer from the client's end.  The client needs
two \rdma\ updates to perform an append -- first to write (and
persist) a new log record at the log tail, and second to write (and
persist) the tail pointer reflecting the new tail of the log.  This
compound update provides the compound \rdma\ update use case in our
experiments.

In both cases, the server asynchronously garbage collects log records
that have been applied at the server end.

\subsection{Experimental Setup}

Our experiments were conducted in a single client and single server
setting, where both systems hosted a dual-socket
Intel\textregistered\ Xeon\textregistered\ E5-2600 processors with 8
hyperthreaded cores per socket with a total of 48 GB of memory.  
The systems run the Fedora 25 distribution of Linux. We
emulate \pmem\ with DRAM.  Each system contains a Mellanox ConnectX-4
100 Gb/s InfiniBand 
RNIC that is used to communicate over the \rdma\ network fabric.  The
client and server communicate with each other via a Mellanox SB7700
36 port 100 Gb/s InfiniBand switch.
Our underlying RDMA framework used busy-waiting for completions rather 
than sleeping while awaiting a completion event.

In our experiments, we emulated a \rdma\ \flush\ with a \rdma\ \Read.
We cannot correctly emulate a non-posted \rdma\ \Write.  However, for
performance estimation of a \rdma\ \flush\ followed by a non-posted
\rdma\ \Write, we can use a \rdma\ \Read\ followed by pipelined
\rdma\ \Write\ and a second \rdma\ \Read.  The \rdma\ \Write\ can be
ordered before the first \rdma\ \Read\ at the server.  However, the
second \rdma\ \Read will not be reordered with the first \rdma\ \Read,
and we believe will give a reasonable estimate for the overhead of a
non-posted \rdma \Write, although it does not enforce the correct
ordering semantics.

\begin{figure*}[t]
  \centering
  \input{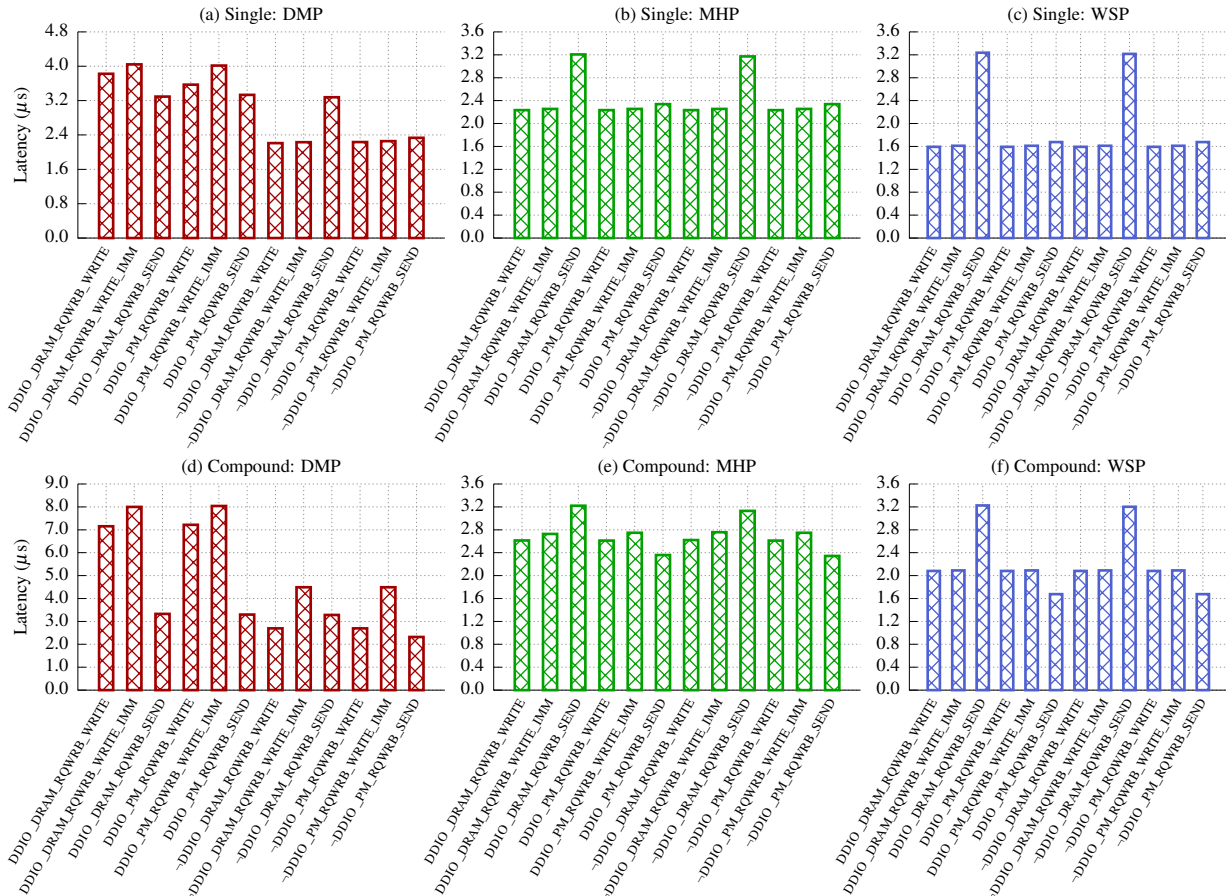}
  \caption{Latencies of remote persistence of \rlog\ appends for
    singleton and compound \rdma\ update based implementations.  In
    all cases, the client appends a 64-byte wide log record.}
\label{fig:latencies}
\end{figure*}

\subsection{Singleton \rdma\ Updates}

\autoref{fig:latencies} (a), (b), and (c) show the latencies of
\rlog\ append operations for the various server configurations.  The
various methods for remote persistence indeed have a significant
impact on latency of appends.  The more general trend is toward a
sizable difference between one-sided and two-sided (classic message
passing) operations, where the former outperforms the latter by up to
$50\%$.

The persistence domains also have a significant impact on append
latencies: When transitioning from DMP to MHP, the observed difference
largely reflects the difference in method of remote persistence.  For
instance, for the DDIO\_DRAM\_RQWRB\_WRITE bars, MHP performs
significantly better than DMP since the former uses one-sided
\rdma\ operations compared to the two-sided operations used by the
latter, which lead to a ping-pong of messages between the client
(requester) and server (responder) -- a full round trip with
additional CPU processing on the server's end to flush cache target
lines.  For both, MHP and WSP, the foundational difference between
performance boils down to whether the \rdma\ update is performed (and
persisted) using one-sided operations or message passing based
ping-pong between the client and the server, with the latter incurring
the round-trip overhead for messages sent back and forth between the
client and server.  We also note that \rdma\ \flush\ has a significant
impact on latency.  In WSP, omission of \rdma\ \flush\ in a one-sided
\rdma\ update drops its latency to $1.6$ microseconds (a $25\%$
reduction in latency from the one-sided \rdma\ updates in MHP).
Overall, as expected, WSP enables the best latency for remote
persistence using \rdma\ operations.

\ddio\ appears to selectively have a negative impact on performance of
some configurations in DMP, particularly for \rdma\ \Write\ and \rdma
\WriteID, when the \RQWRB\ is placed in \pmem.  But this effect is
indirect in that it forces a two-sided operation to ensure that the
remotely updated cache lines are flushed to the DMP domain.  Placement
of \RQWRB\ has a significant performance indirectly as well in that if
the \RQWRB\ is placed in \pmem, \rdma\ \Send\ can
be treated by the client as a purely one-sided operation, and hence
gain the performance advantage of one-sided operations.  However, care
must be taken by programmers on balancing consumption of receive queue
buffers at the server end with the rate of \rdma\ \Send\ and
\rdma\ \WriteID\ operations coming from the client.  Each such
operation consumes a receive queue buffer on the server's end, and the
server must quickly recycle these buffers in order to continue
receiving messages from the client.  If the server is too slow,
resource availability timeouts may be triggered on the client's end
leading to performance jitter.

\subsection{Compound \rdma\ Updates}

Compound update latency results appear in~\autoref{fig:latencies} (d),
(e), and (f).  As in the case of singleton updates, the server
configuration and resulting method of remote persistence has a
significant impact on latencies of remote persistence.  The universal
message passing based approach, which had a negative performance
impact for singleton updates, appears to have a significant advantage
in servers supporting the DMP domain.  The advantage is that the two
updates -- the log tail record and the tail pointer -- can be packaged
in a single message by the client, which keeps the operation to a
single round trip.  In contrast, use of \rdma\ \Write\ and
\WriteID\ with message passing leads to two round trips leading to
more than $2X$ latency in DMP when \ddio\ is turned on.  However, MHP
unlocks the capability of doing one-sided compound \rdma\ updates lead
to significant latency improvements in \rdma\ \Write\ and
\WriteID\ based methods, which end up performing with a latency up to
$20\%$ better than the latency of message passing.  This gain is more
pronounced to $30\%$ for WSP.

Similar to singleton updates, \ddio\ appears to have an indirect
negative performance impact on DMP configurations for \rdma\ updates
done using \rdma\ \Write\ and \rdma\ \WriteID, in that it forces
additional message passing (and cache line flushing at the server)
overheads for remote persistence.  The non-posted \rdma\ \Write\ based
method is enabled when \ddio\ is turned off, and appears to deliver a
big performance improvement.  In general, turning \ddio\ off enables
compound updates using \rdma\ \Write\ and \WriteID\ to be done using
just one-sided \rdma\ operations, where the big performance manifests.
Notice however, that the latency of \rdma\ \WriteID\ does not drop as
much.  This is because non-posted writes enable pipelining of updates
and \rdma\ \flush.  However since there is no non-posted version of
\rdma\ \WriteID\ available, the \rdma\ \WriteID\ based method, incurs
overhead of completion of the first \rdma\ \flush\ before issuing the
second \rdma\ \WriteID.  As expected, \ddio\ has no effect on MHP and
WSP configurations.

Placement of \RQWRB\ in \pmem\ enables a big optimization in the
\rdma\ \Send\ based method in that, provided the persistence domain is
either MHP or WSP, or DDIO is turned off, \rdma\ \Send\ can be treated
by the client as a one-sided \rdma\ operation.  
As in the case of
singleton updates, absence of \rdma\ \flush{}es in servers configured
to support WSP boosts latency of remote persistence by close to
$20\%$.

\section{Conclusions}

We presented the first comprehensive taxonomy of methods for
persistence of \rdma\ updates to remote persistent memory.  Our
taxonomy spans server configurations along three different axes: (i)
the persistent domain of the system, (ii) use of \ddio, and
(iii) placement of \RQWRB{}s in DRAM or \pmem.  We showed how these
configurations affect the methods to correctly and efficiently enforce
persistence of \rdma\ updates.  We also included some of the recent
advances in the IBTA standards community~\cite{burstein18,grun18} in
our analysis.  Our detailed analysis covered persistence of singleton
\rdma\ updates as well as compound \rdma\ updates that need to be
persisted in the order they were issued from the requester.  We find
that the methods to correctly, and efficiently, persist \rdma\ updates
vary significantly based on the underlying system's configuration
parameters enumerated above.  Programmers must be extremely careful in
applying these methods -- application of the wrong method can lead to
significant performance overheads, and even critical data
inconsistencies in the face of system failures.

Our evaluation demonstrated several interesting performance trade offs
between available methods for persistence of \rdma\ updates.  In
particular, we find that remote persistence done using one-sided
\rdma\ operations (\rdma\ \Write, \rdma\ \WriteID, \rdma\ \flush, and
even \rdma\ \Send\ in cases where the \RQWRB\ resides in \pmem) is
generally more efficient than remote persistence enforced using
\rdma\ \Send\ based message passing.  In the end, we believe the
workloads requirements may determine the best choices.  A
client may need to perform a complex set of non-contiguous updates at
the server, which would be better served by a single
\rdma\ \Send\ based remote procedure call (RPC)~\cite{kalia16}.  

The newly proposed non-posted \rdma\ \Write\ based method is also
quite effective in delivering better performance.  However, this
\rdma\ extension seems useful only in a small part of the space of
system configurations we explored in the paper.  Perhaps, that small
part itself could represent the dominant system configurations used in
the industry.  It remains to be seen what configurations will be used
widely in the future.

Given the wide range of choices of remote persistence, it may be
reasonable to build a single \rdma\ library that transparently applies
the correct method of remote persistence for a given system and
application.  There may be interesting subtleties that may lead to
sub-optimal performance, and even correctness issues in the face of
failure.  However, we leave the exploration for future work.  Another
interesting aspect that remains to be explored is implications of
these choices for remote persistence on memory persistency
models~\cite{kolli17,pelley14}.

\balance
{\normalsize \bibliographystyle{acm}
\bibliography{main}}


\end{document}